\documentclass[twocolumn,showpacs,preprintnumbers,amsmath,amssymb]{revtex4}

\newcommand{\bib}{\bibitem}
\newcommand{\bea}{\begin{eqnarray}}
\newcommand{\eea}{\end{eqnarray}}
\newcommand{\beq}{\begin{equation}}
\newcommand{\eeq}{\end{equation}}
\newcommand{\non}{\nonumber}
\newcommand{\noi}{\noindent}
\newcommand{\da}{\dagger}
\newcommand{\al}{\alpha}

\newcommand{\ga}{\gamma}
\newcommand{\de}{\delta}

\newcommand{\ep}{\epsilon}

\newcommand{\om}{\omega}

\newcommand{\ta}{\theta}
\newcommand{\pa}{\partial}
\newcommand{\ra}{\rangle}
\newcommand{\la}{\langle}
\newcommand{\ct}{\cite}

\usepackage{graphicx,epsfig} 
\usepackage{dcolumn} 
\usepackage{bm} 

\begin{document}

\title{Nonadiabatic charge pumping by oscillating potentials in 
one dimension: \\ results for infinite system and finite ring}
\author{Abhiram Soori and Diptiman Sen}
\affiliation{Center for High Energy Physics, Indian Institute of Science,
Bangalore 560012, India}
\date{\today}

\begin{abstract}
We study charge pumping when a combination of static potentials and 
potentials oscillating with a time period $T$ is applied in a one-dimensional 
system of non-interacting electrons. We consider both an infinite system using
the Dirac equation in the continuum approximation, and a periodic ring with a 
finite number of sites using the tight-binding model. The infinite system 
is taken to be coupled to reservoirs on the two sides which are at the same 
chemical potential and temperature. We consider a model in which oscillating 
potentials help the electrons to access a transmission resonance produced by 
the static potentials, and show that non-adiabatic pumping violates the simple
$\sin \phi$ rule which is obeyed by adiabatic two-site pumping. For the ring, 
we do not introduce any reservoirs, and we present a method for 
calculating the current averaged over an infinite time using the time 
evolution operator $U(T)$ assuming a purely Hamiltonian evolution.
We analytically show that the averaged current is zero if the Hamiltonian 
is real and time reversal invariant. Numerical studies indicate another 
interesting result, namely, that the integrated current is zero for {\it any} 
time-dependence of the potential if it is applied to only one site. Finally 
we study the effects of pumping at two sites on a ring at resonant and 
non-resonant frequencies, and show that the pumped current has different 
dependences on the pumping amplitude in the two cases.
\end{abstract}

\pacs{73.23.-b, 73.63.Nm, 72.10.Bg}
\maketitle

\section{Introduction}

The idea that oscillating potentials applied to certain points in a 
one-dimensional system can pump a net charge between two reservoirs at the 
same chemical potential has been studied extensively for many years, both 
theoretically [1-42] and experimentally 
\ct{switkes,taly1,cunningham,taly2,leek,dicarlo,kaestner}. For the case of 
non-interacting electrons, theoretical studies of this phenomenon have used 
adiabatic scattering theory 
\ct{avron1,avron2,entin1,entin2,moskalets3,cohen,graf,braun2,hwang}, Floquet
scattering theory \ct{kim,moskalets1,moskalets2}, the non-equilibrium Green 
function formalism \ct{wang,arrachea1,arrachea2,torres}, and the equation of 
motion 
approach \ct{agarwal1,agarwal2}. The case of interacting electrons has been 
studied using a renormalization group method for weak interactions \ct{das},
and the method of bosonization for arbitrary interactions [51-61]. Other 
interesting studies of pumping include adiabatic quantum pumping in graphene 
where the electrons obey the Dirac equation \ct{prada}, and classical pumping 
on a finite ring by oscillating hopping rates at two sites \ct{jain}.

With the exception of a few papers 
\ct{moskalets1,wang,arrachea1,arrachea2,torres,kaestner}, the earlier studies 
of charge pumping have generally considered systems in which oscillating 
potentials are applied to two or more sites. In such cases, it is known that 
if the oscillation frequency $\om$ is small, the dc part of the pumped current
is proportional to $\om$; the charge pumped per cycle (with time period $2\pi /
\om$) therefore has a finite value in the adiabatic limit $\om \to 0$. However,
it has been noted in Refs. \onlinecite{moskalets1} and 
\onlinecite{wang,arrachea1,arrachea2,torres} that an oscillating potential 
applied to a single site can also pump charge provided that the system has 
no left-right symmetry; 
this can happen if, for instance, appropriate static potentials are present.
Most studies of charge pumping have also been limited to infinite systems in 
which the left and right sides of the system (called the leads or reservoirs)
are associated with certain chemical potentials and temperatures. Charge 
pumping on a finite ring has been studied in a few papers for adiabatic 
\cite{moskalets3,cohen,graf,braun2} and nonadiabatic situations 
\cite{moskalets2,arrachea1,arrachea2}.

In this paper, we will study charge pumping in both an infinite system as well
as on a finite ring for non-interacting electrons. We will study the effects 
of oscillating potentials applied to either one site or more than one site. We
will not assume the oscillation frequency to be small (i.e., the adiabatic 
limit), but will assume the oscillation amplitudes to be small. For the 
infinite system, we will assume the existence of reservoirs, but for the 
finite ring, we will assume that the system is not coupled to any reservoirs 
and that the time evolution is purely Hamiltonian. We will see that results 
obtained in the two cases differ in some interesting ways. We will restrict 
our analysis to zero temperature and spinless electrons, since spin does not 
play an essential role in the absence of interactions between the electrons.

The plan of the paper is as follows. In Sec. II, we will examine charge 
pumping in the infinite system. For convenience, this will be studied in the 
continuum limit using a linearized form of the energy-momentum relation; 
namely, we will use the massless Dirac equation with both right and left 
moving modes. We will assume the system to be coupled to reservoirs on the 
two sides, and the chemical potentials and temperature of these will be 
introduced using the formalism of Refs. \onlinecite{moskalets1} and 
\onlinecite{moskalets2}. This implicitly assumes that an electron, after 
passing through the region with the static and oscillating potentials, 
equilibrates with whichever lead it enters; the precise mechanism for energy 
or momentum relaxation in the reservoirs will not be specified in our 
calculation. As a 
specific example, we will consider a model in which the static potentials 
have transmission resonances at certain energies, and potentials oscillating 
at exactly the right frequency can then lead to enhanced charge pumping 
\cite{wiel,kohler}; in this model, we show how a competition between different
processes can lead to either maxima or minima in the pumped current. We find 
that the pumped charge is of second order in the strengths of the oscillating 
potentials and can be much larger for two-site pumping compared to one-site 
pumping. We also find that the `sin $\phi$' rule, which has been discussed 
earlier in the adiabatic limit $\om \to 0$ for pumping by two oscillating 
potentials with a phase difference of $\phi$ \cite{brouwer1,switkes}, breaks 
down if $\om$ is larger than the resonance width.

In Sec. III, we will study charge pumping on a finite ring using a 
tight-binding Hamiltonian to describe the electrons; no reservoirs will
be introduced. We will present a formalism for calculating the long-time 
averaged current if there are potentials oscillating with a time period $T$; 
this is done by considering the eigenstates of the unitary time evolution 
operator $U(T)$. Next, we will assume that the system begins in a specific 
initial state corresponding to a particular filling at zero temperature. We 
will then evolve the system using only the Hamiltonian, without introducing 
any mechanisms for momentum relaxation and phase decoherence; this means that 
the system will never reach a steady state and that its properties depend on 
the initial state. We will show analytically that if the Hamiltonian is real 
and time reversal invariant, then charge pumping cannot occur even if the 
system has no left-right symmetry. Numerically, we also find that if the 
Hamiltonian is real and if an oscillating potential is applied to only one 
site, then charge pumping does not occur even if the oscillating potential is 
not time reversal invariant and the system has no left-right symmetry. It is 
therefore necessary to apply oscillating potentials to at least two sites in 
order to pump charge. We will also study what happens at both non-resonant 
and resonant frequencies; the latter means that the oscillation 
frequency is equal to the energy difference between a filled state and an 
empty state of the time-independent part of the Hamiltonian. (We would like 
to mention here that charge pumping on a ring at resonant and non-resonant 
frequencies has also been studied earlier in Ref. \onlinecite{moskalets2}). 
In the Appendix, we will present details of the calculation of the eigenstates
of $U(T)$ to first order in the oscillating potentials, both for the 
non-resonant and resonant cases. We will show there that the pumped 
charge can receive contributions at either zero-th or first order in the 
oscillating potential in the resonant case but only at second order in the 
non-resonant case. In Sec. IV, we will summarize our results and emphasize
the different assumptions that we have made in the models defined on the 
infinite line and on a ring. Finally, we will discuss possibilities for 
experimentally testing our results in two kinds of systems, namely, 
mesoscopic rings and aromatic molecules.

\section{Charge pumping on an infinite line}

In this section, we will study charge pumping on an infinite line. Our model
consists of a gapless system of non-interacting spinless electrons subject 
to some static and oscillating potentials. We will assume that the regions 
lying far to the left and far to the right of all the potentials have the 
same chemical potential given by the Fermi energy $E_F$ (we will work at 
zero temperature). Assuming that the oscillating frequency is small compared
to the bandwidth, we will linearize the energy-momentum dispersion around
$E_F$. Let us denote the Fermi velocity and Fermi wave number by $v_F = 
(dE/dp)_{E=E_F}$ and $k_F$ respectively. [The relation between $k_F$ and
$E_F$ is governed by the underlying microscopic model. For instance, in
a tight-binding lattice model with the dispersion $E=-2 \ga \cos k$,
where $\ga$ is the nearest-neighbor hopping amplitude,
we have $E_F =-2 \ga \cos k_F$ where $0 < k_F < \pi$]. 
We can then define the electron field operator
\beq \psi (x) ~=~ \psi_R (x) ~e^{ik_F x} ~+~ \psi_L (x) ~e^{-ik_F x}, 
\label{psi} \eeq
where $\psi_L$ and $\psi_R$ are the fermionic field operators for the left and
right moving electrons. 
In terms of these fields, the Hamiltonian in the presence of several 
point-like potentials is given by $H = H_0 + V$, where
\bea H_0 &=& \int dx ~i v_F ~(- ~\psi_R^\da \frac{\pa \psi_R}{\pa x} ~+~ 
\psi^\da_L \frac{\pa \psi_L}{\pa x}) , \non \\
V &=& \int dx ~\sum_p ~\de (x-x_p) ~U_p(t) ~\psi^\da (x) \psi (x). 
\label{ham1} \eea
In the absence of the potential $V$, the eigenstates of $H$
are given by $\exp [i(\pm kx-Et)]$, where $E=v_F k$, and $\pm k$ refer to
right and left moving modes respectively; note that we are measuring 
the energy $E$ and the wave number $k$ with respect to $E_F$ and $k_F$
respectively, and we have set $\hbar = 1$. 
Using Eq. (\ref{psi}), we find that the equations of motion for $\psi_R$ 
and $\psi_L$ are given by \ct{agarwal3}
\bea & & i~ \frac{\pa \psi_R}{\pa t} ~+~ i v_F ~ \frac{\pa \psi_R}{\pa x} 
\non \\
& & =~ \sum_p \de (x-x_p) ~U_p (t) ~(\psi_R ~+~ \psi_L e^{-i2k_F x_p}), \non \\
& & i~ \frac{\pa \psi_L}{\pa t} ~-~ i v_F ~\frac{\pa \psi_L}{\pa x} \non \\
& & =~ \sum_p ~\de (x-x_p) ~U_p (t) ~(\psi_L ~+~ \psi_R e^{i2k_F x_p}). 
\label{eom} \eea
We can solve these equations by integrating over small regions from 
$x_p - \ep$ to $x_p + \ep$
to find the discontinuities in the fields at $x=x_p$.

Let us now consider two types of point-like potentials: $a_m \de (x - x_m)$ 
which are time independent, and $w_n (t) \de (x - x_n)$ which oscillate in time
as $w_n (t) = b_n \cos (\om t + \phi_n)$. We will assume that the oscillation
amplitudes $b_n$ are small compared to $v_F$ and will generally calculate the 
pumped charge to the lowest non-zero order in the $b_n$'s. Let us first assume
that $b_n = 0$ for all $n$, and that we can completely solve the problem with 
the static potentials. This can be described in terms of a scattering matrix 
$S$ as follows. If $\cal R$ denotes the region within which all the static 
potentials are present, then an electron incoming from a region far to 
the left of $\cal R$ (which we denote as $x \ll {\cal R}$) with unit
amplitude and energy and wave number given by $E$ and $k=E/v_F$ respectively 
will have a wave function $\psi (x) e^{-iEt}$ given by
\bea \psi (x) &=& e^{ikx} ~+~ r_L (E) ~e^{-ikx} ~~{\rm for} ~~x \ll {\cal R}, 
\non \\ 
&=& t_R (E) ~e^{ikx} ~~{\rm for} ~~x \gg {\cal R}, \eea
where $r_L$ and $t_R$ denote the reflection and transmission amplitudes. 
Similarly, an electron 
incoming from a region far to the right of $\cal R$ with unit amplitude and 
energy and wave number given by $E$ and $-k = -E/v_F$ respectively has the 
wave function $\psi (x) e^{-iEt}$, where
\bea \psi (x) &=& e^{-ikx} ~+~ r_R (E) ~e^{ikx} ~~{\rm for} ~~x \gg {\cal R}, 
\non \\
&=& t_L (E) ~e^{-ikx} ~~{\rm for} ~~x \ll {\cal R}. \eea
Assuming that all these reflection and transmission amplitudes are known,
let us now turn on the oscillating potentials, all of which we take to lie
to the left of $\cal R$, and proceed as follows.

An electron incoming from the left of $\cal R$ with unit amplitude and
energy (wave number) $E_0$ ($k_0 = E_0/v_F$) will now have a wave function 
$\psi (x,t)$ of the form 
\bea \psi &=& e^{i(k_0 x - E_0 t)} ~+~ \sum_j ~r_{L,j} ~e^{i(-k_j x - E_j 
t)} ~~{\rm for} ~~x \ll {\cal R}, \non \\
&=& \sum_j ~t_{R,j} e^{i(k_j x - E_j t)} ~~{\rm for} ~~x \gg {\cal R}, 
\label{inleft} \eea
where $E_j = E_0 + j \om$, $k_j = E_j/v_F$, and $r_{L,j}$ and $t_{R,j}$
will now generally be functions of both $E_0$ and $E_j$. Similarly, an 
electron incoming from the right of $\cal R$ with unit amplitude and
energy (wave number) $E_0$ ($k_0 = E_0/v_F$) will have a wave function 
$\psi (x,t)$ of the form 
\bea \psi &=& e^{i(-k_0 x - E_0 t)} ~+~ \sum_j ~r_{R,j} ~e^{i(k_j x - E_j 
t)} ~~{\rm for} ~~x \gg {\cal R}, \non \\
&=& \sum_j ~t_{L,j} e^{i(-k_j x - E_j t)} ~~{\rm for} ~~x \ll {\cal R}, 
\label{inright} \eea
where $E_j = E_0 + j \om$, $k_j = E_j/v_F$, and $r_{R,j}$ and $t_{L,j}$ will 
be functions of both $E_0$ and $E_j$. 

The sums over the side band index $j$ 
in Eqs. (\ref{inleft}-\ref{inright}) will go from $-\infty$ to $\infty$ for a 
Dirac electron. But, in practice, $j$ gets cut off for two reasons. First, if
the energy $E_j$ goes above or below the bandwidth of the system, the
corresponding wave function decays exponentially as $|x| \to \infty$;
such a state does not carry any current. Secondly, if all the pumping 
amplitudes $b_n$ are small, one can show, using Eqs. (\ref{eom}) recursively,
that the leading order terms in $r_{R/L,j}$ and $t_{R/L,j}$ are given by 
$|b_n|^{|j|}$ if $j \ne 0$; this goes to zero exponentially as $|j| \to 
\infty$. Hence, if $\om$ is much smaller than the bandwidth and $E_0$ is not 
too close to the edges of the band, the contributions of the higher side 
bands become very small long before the energy $E_j$ reaches the band edges.

We will now calculate the different reflection and transmission 
amplitudes using Eqs. (\ref{eom}). To first order in $b_n$, only the first 
side bands with $j = \pm 1$ survive. If the oscillating potentials are given by
\beq \sum_n ~\de (x-x_n) ~b_n ~\cos (\om t + \phi_n), \eeq
we find that
\bea t_{R,\pm 1} &=& - ~\frac{i}{2v_F} ~t_R (E_{\pm 1}) ~\sum_n ~b_n e^{\mp i
\phi_n} \non \\
& & \times~ [~e^{\mp i \om x_n/v_F} ~+~ r_L (E_0) e^{-i(k_{\pm 1}+k_0+2k_F)
x_n}], \non \\
r_{R,\pm 1} &=& - ~\frac{i}{2v_F} ~t_R (E_{\pm 1}) ~t_L (E_0) \non \\
& & \times ~\sum_n ~b_n e^{\mp i \phi_n} ~e^{-i(k_{\pm 1}+k_0+2 k_F) x_n}.
\label{trr} \eea
Similar expressions can be derived for $t_{L,\pm 1}$ and $r_{L,\pm 1}$,
but these will not be required below. 
The expressions in Eqs. (\ref{trr}) can be understood as arising from a 
sum over paths as explained in Ref. \cite{agarwal2}.

Given all the transmission and reflection amplitudes $t_{R/L,j}$ and 
$r_{R/L,j}$, the dc part of the current in, say, the right lead is given by
\ct{moskalets1,moskalets2}
\bea I_{R,dc} &=& q ~\int_{-\infty}^\infty \frac{dE_0}{2\pi} ~\sum_j ~[~
|r_{R,j}|^2 ~\{f_R (E_0) - f_R (E_j)\} \non \\
& & \quad \quad \quad \quad \quad \quad ~+~ |t_{R,j}|^2 ~\{f_L (E_0) - f_R
(E_j)\} ~], \non \\
& & \label{irdc1} \eea
where $q$ is the charge of the electron, and $f_\al (E)=1/[e^{\beta (E-
\mu_\al)} + 1]$ is the Fermi function in the lead $\al$. At zero
temperature, $f_\al (E) = 1$ if $E < \mu_\al$ and 0 if $E > \mu_\al$.
If we assume zero bias, $\mu_R = \mu_L = 0$ (we are defining energy with 
respect to $E_F$), and restrict ourselves to terms of second order in the 
$b_n$'s, we obtain the expression 
\bea I_{R,dc} & & = ~q ~\int_{-\om}^0 ~\frac{dE_0}{2\pi} ~[|r_{R,1}|^2 ~+~ 
|t_{R,1}|^2] \non \\
& & ~- q ~\int_0^{\om} ~\frac{dE_0}{2\pi} ~[|r_{R,-1}|^2 ~+~ |t_{R,-1}|^2].
\label{irdc2} \eea
In the adiabatic limit $\om \to 0$, we can show from this that, up to order 
$\om$, the current will involve only cross-terms of the form $\sum_{m<n} b_m 
b_n \sin (\phi_m - \phi_n)$ multiplied by terms involving $x_m$ and $x_n$.
Let us call this the `$\sin \phi$' rule; it implies that two potentials cannot
pump current if their phase difference is 0 or $\pi$. We also see that if 
there is an oscillating potential at only site, there is no current at order 
$\om$. These features arise because of the near cancellation in Eq. 
(\ref{irdc2}) between $|r_{R,1}|^2$ and $|r_{R,-1}|^2$ and between 
$|t_{R,1}|^2$ and $|t_{R,-1}|^2$ at order $\om$. However, the $\sin \phi$ rule
does not hold, and an oscillating potential at even one site can pump current,
if $\om$ is larger than the resonance width; in that case $E_0 + \om$ and 
$E_0 - \om$ will be sufficiently different from each other so that
$|r_{R,1}|^2$ and $|r_{R,-1}|^2$ or $|t_{R,1}|^2$ and $|t_{R,-1}|^2$ are 
no longer almost equal to each other. The violation of the $\sin \phi$ rule
will be explicitly demonstrated below for the case of pumping at two
sites.
 
We will now illustrate the above results for the case in which there are
two static $\de$-function potentials given by
\beq a_1 \de (x - x_1) + a_2 \de (x - x_2). \eeq
Defining $u_i = a_i/v_F$, we find the following expressions for the
reflection and transmission amplitudes as functions of the wave number $k$,
\bea t_R &=& t_L = \frac{1}{1+i(u_1+u_2)-u_1 u_2 (1-e^{i2(k+k_F)(x_2-x_1)})},
\non \\
r_L &=& t_R ~[-i u_1e^{i 2(k+k_F)x_1} -i u_2e^{i 2(k+k_F)x_2} \non \\
& & ~~~~+ u_1u_2 (e^{i 2(k+k_F)x_1} -e^{i 2(k+k_F)x_2})], \non \\
r_R &=& t_R ~[-i u_1e^{-i 2(k+k_F)x_1} - i u_2e^{-i 2(k+k_F)x_2} \non \\
& & ~~~~- u_1u_2 (e^{-i 2(k+k_F)x_1} -e^{-i 2(k+k_F)x_2})]. \label{tr} \eea
These expression simplify considerably if $u_1 = u_2 = u$, i.e., if we have 
a symmetric double barrier. Defining $\ta = (k+k_F)(x_2 - x_1)$, we obtain
\beq |t_R|^2 ~=~ |t_L|^2 ~=~ \frac{1}{1 ~+~ 4u^2 ~(\cos \ta ~+~ u \sin 
\ta)^2}. \eeq
We then see that $|t_R| = |t_L| = 1$, i.e., there is a resonance, whenever
$\tan \ta = -1/u$. This occurs at $\ta = (n+1/2) \pi$ for $u \to 0$ and at
$n\pi$ for $u \to \infty$. For $|u| \gg 1$, the resonances are very sharp,
with $|t_R|^2$ dropping to $1/2$ when one deviates from one of the resonant
values of $\ta$ by $1/(2u^2)$, namely, when $k_F$ deviates from one of the 
resonant values by $1/(2u^2 |x_2 - x_1|)$. 

Similar to other models considered in the earlier literature,
we can now study what happens when an electron is incident on the symmetric
double barrier with an energy $E_0$ which is {\it not} at resonance, but the
pumping frequency is such that $E_0 + \om$ or $E_0 - \om$ is equal to one of 
the resonant energies called $E_r$. Then the oscillating potentials can
change the energy of the electron to $E_r$ which can then transmit through
the double barrier. Thus the pumping can help the electrons to access a 
resonance.
We will compare below the cases of pumping at one site versus two sites, and
also study the dependence of the pumped current on the phase difference 
$\phi$ in the case of two-site pumping.

\begin{figure}[htb]
\begin{center} \epsfig{figure=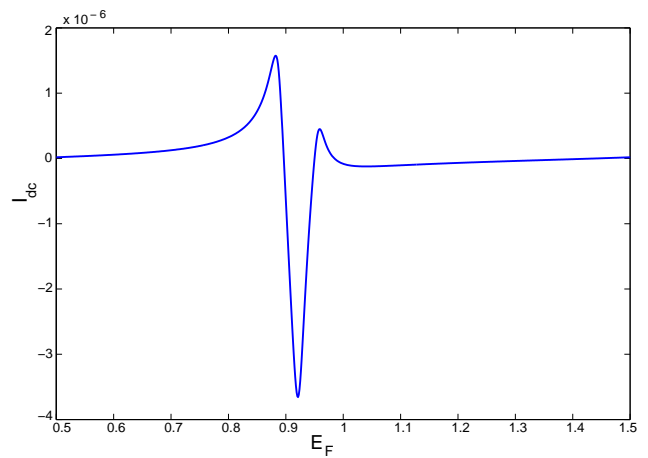,width=8.6cm} \end{center}
\caption{(Color online) Current pumped through a symmetric double barrier by 
an oscillating potential applied at one site as a function of $E_F$. The 
barrier strengths $u_i$ are both equal to 4, and their positions are at 0 and 
$\pi$. The resonant energy is $E_r = 0.917$, and no bias is applied. The 
oscillating potential is applied at the position $-\pi$ with an amplitude 
$0.2$ and frequency $0.035$. We set $v_F = 1$. The locations of the two 
maxima and the minimum in the middle are discussed in the text.} \end{figure}

In Fig. 1, we show the pumped dc current $I_{R,dc}$ versus the Fermi energy 
$E_F$ when there is a static and symmetric double barrier and an oscillating 
potential is applied at one site lying on the left of the double barrier.
As $E_F$ increases, we observe first a maximum at
$E_F = 0.883$, then a minimum at $0.921$, and then a small maximum at $0.960$.
These features arise for the followings reasons. The first maximum at $0.883$
coincides with $E_r - \om$ within a spread given by the resonance width 
$1/(2\pi u^2) \simeq 0.010$ (note that this is significantly less than the
pumping frequency $\om = 0.035$). This occurs because an electron approaching
from the left reservoir with an energy equal to $E_F = E_r - \om$ can get 
boosted up to the energy $E_r$ due to the oscillating potential; it can then 
transmit through the double barrier. 
The minimum at $0.921$ coincides with $E_r$; note that this corresponds to a 
negative current, namely, the current is flowing to the left. This occurs 
due to a combination of two effects. An electron approaching from the right 
reservoir with an energy equal to $E_F = E_r$ transmits through
the double barrier; it can then get boosted to the energies $E_r \pm
\om$ by the oscillating potential, which means that it cannot transmit back 
to the right through the double barrier. Similarly, an electron approaching 
from the left reservoir with the an energy equal to $E_F = E_r$ can get
boosted to the energies $E_r \pm \om$ by the oscillating potential, which 
means that it cannot transmit to the right through the double barrier. Hence,
in both cases, the electron finds it easier to go to the left than to the
right, leading to a net current to the left. Finally, the small maximum at
$0.960$ coincides with $E_r + \om$. This maximum is not very robust; its height
can change easily depending on the values of the various parameters because
it is a result of several competing processes. An electron approaching from 
the right reservoir with an energy equal to $E_r$ (which is less than $E_F$)
transmits through the double barrier; it can then get boosted to the 
energies $E_r \pm \om$ by the oscillating potential and then 
escape to the left reservoir. On the other hand, an electron approaching 
from the left reservoir with an energy equal to $E_r \pm \om$ can get 
boosted to the energy $E_r$ by the oscillating potential, which means that 
it can then transmit to the right through the double barrier. Depending on 
which of these is larger, the net current can be positive or negative.

\begin{figure}[htb]
\begin{center} \epsfig{figure=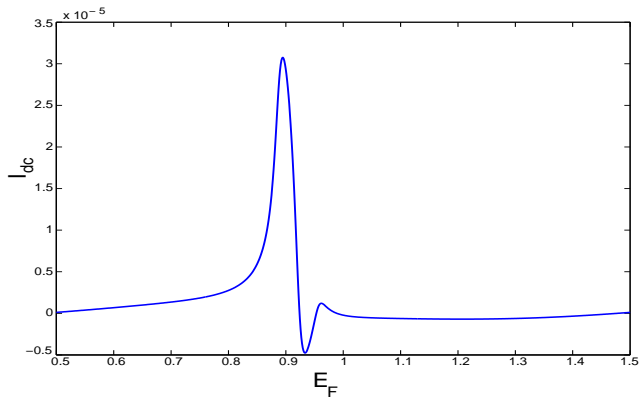,width=8.6cm,height=5.4cm} \end{center}
\caption{(Color online) Current pumped through a symmetric double barrier 
by oscillating potentials applied at two sites.
The barrier strengths are both equal to 4, and their positions are at 0 and 
$\pi$. The resonant energy is $E_r = 0.917$, and no bias is applied. 
The oscillating potentials are applied at the positions $-\pi$ and $-2\pi$
with the same amplitude $0.2$ and frequency $0.035$, but with a phase 
difference of $\pi/2$. We set $v_F = 1$. The locations of the maxima and 
the minimum are discussed in the text.} \end{figure}

In Fig. 2, we show the pumped dc current versus the Fermi energy 
$E_F$ when there is a symmetric double barrier and oscillating potentials
are applied at two sites (lying on the left of the double barrier)
with a phase difference of $\pi/2$ between the two potentials. Just as in
Fig. 1, as $E_F$ increases, we observe first a maximum at $E_F = 0.895$, 
then a minimum at $0.932$, and then a small maximum at $0.963$. The reasons 
for all these features are the same as the ones discussed above for Fig. 1. 
We observe that the maximum value of the pumped current in Fig. 2 is
about 10 times the corresponding value in Fig. 1, showing that pumping by
two oscillating potentials can be much more effective than by one oscillating
potential. This can be qualitatively understood as follows. If oscillating
potentials are applied to $K$ sites, and their effects add up constructively,
the reflection and transmission amplitudes in the first side band would be of
the order of $K$ times what one would get if there was pumping at only one 
site; this is evident from Eqs. (\ref{trr}). The pumped current would 
therefore be magnified by a factor of $K^2$.

\begin{figure}[htb]
\begin{center} \epsfig{figure=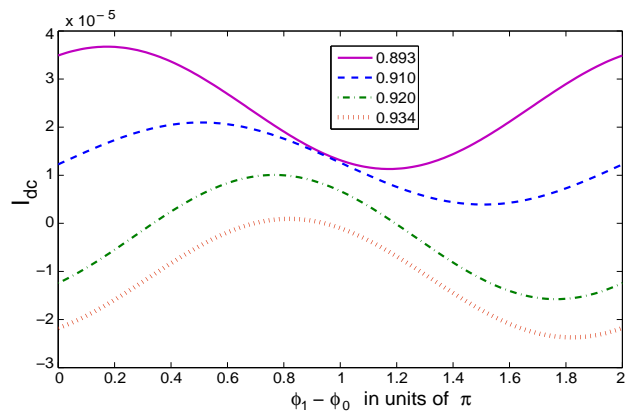,width=8.6cm,height=5.5cm} \end{center}
\caption{(Color online) Current pumped through a symmetric double barrier by 
oscillating potentials applied at two sites as a function of the phase 
difference $\phi_2 - \phi_1$, for
four different Fermi energies given by $0.893$ (magenta solid line), $0.910$
(blue dashed line), $0.920$ (green dotted-dashed line) and $0.934$ (red dotted
line). The barrier strengths are both equal to 4, and their positions are at 0
and $\pi$. The resonant energy is $E_r = 0.917$, and no bias is applied. The 
oscillating potentials are applied at the positions $-\pi$ and $-2\pi$ with 
the same amplitude $0.2$ and frequency $0.035$. We set $v_F = 1$.} \end{figure}

In Fig. 3, we show the pumped dc current as a function of the phase difference
between oscillating potentials applied at two sites, for four different values
of the Fermi energy $E_F$. These values have been chosen to lie between the
first maximum and the minimum observed in Fig. 2. We observe that all the
four curves are approximately of the form $d_1 + d_2 \sin (\phi_2 - \phi_1 + 
d_3)$, where the values of $d_1$, $d_2$ and $d_3$ are different for the 
different curves. All of these differ from a simple $\sin \phi$ rule which 
would correspond to the parameters $d_1$ and $d_3$ being equal to 0. This 
violation of the $\sin \phi$ rule occurs here because we have chosen 
$\om$ to be larger than the resonance width. The demonstration of this
violation is one of the main results of this section.

The results in this section can be generalized to finite temperatures
by using the appropriate Fermi functions in Eq. (\ref{irdc1}). One of
the effects of a finite temperature $T$ is to introduce a finite phase
decoherence length given by $l_\phi = \hbar v_F/(k_B T)$ \ct{kane}. We 
expect this to destroy the resonance produced by the static double barrier
when $l_\phi$ becomes smaller than the distance between the barriers;
hence the peaks in the pumped current will become broad and eventually
disappear. 

\section{Charge pumping on a finite ring}

In this section, we will study what happens when oscillating potentials are 
applied to a finite-sized ring which is {\it not} coupled to any reservoirs. 
This study may possibly be of interest in the context of transport in 
mesoscopic rings or even molecular rings \cite{avron2} which are either not 
coupled to any reservoirs or are so weakly coupled to reservoirs that the 
momentum relaxation and phase decoherence times are much longer than the time 
period of the oscillating potentials. (A discussion of the possible 
experimental applications of our work will be presented in Sec. IV). Once 
again, we will consider non-interacting spinless electrons. The Hamiltonian 
will be taken to be of the tight-binding form with nearest-neighbor hopping 
amplitudes and on-site potentials which are either static or oscillating with 
a time period $T$. Namely, we have $H = H_0 + V (t)$, where
\bea H_0 &=& - \sum_{n=1}^N ~\ga ~(c_n^\da c_{n+1} ~+~ c_{n+1}^\da c_n) + 
\sum_{n=1}^N ~a_n ~c_n^\da c_n, \non \\
V(t) &=& \sum_{n=1}^N ~w_n (t) ~c_n^\da c_n, \label{ham2} \eea
where $w_n (t) = w_n (t+T)$ with $T$ denoting the time period. (We will set 
the hopping amplitude $\ga = 1$ in all our numerical calculations). Note that 
we have included all the static potentials in $H_0$. We will impose the 
condition that
\beq \int_0^T ~dt ~w_n (t) ~=~ 0 \eeq
for all $n$; if necessary, this can be ensured by adding a constant
to $w_n (t)$ and subtracting the same constant from $a_n$ in $H_0$.
We are therefore assuming that $V$ satisfies
\beq \int_0^T ~dt ~V (t) ~=~ 0. \label{vzero} \eeq
We will impose periodic boundary conditions, so that $N+1 \equiv 1$ in Eq.
(\ref{ham2}). It will be convenient below to rewrite the operators in
Eq. (\ref{ham2}) as $N \times N$ matrices in the one-particle basis. Namely,
\bea (H_0)_{jk} &=& - ~\ga ~(\de_{j,k+1} + \de_{j,k-1}) ~+~ \sum_n ~a_n ~
\de_{j,k} \de_{j,n}, \non \\
(V(t))_{jk} &=& \sum_n ~w_n (t) ~\de_{j,k} \de_{j,n}. \eea

We will now study various features of this model; in particular, we will
calculate the current averaged over one time period for different choices
of the oscillating potential $V(t)$. We will assume that the system is not 
connected to any external reservoirs and has no mechanism for momentum 
relaxation or phase decoherence. We will begin with a given initial
state in which the lowest $p$ energy levels of the Hamiltonian $H_0$ in
Eq. (\ref{ham2}) are filled (so that the filling fraction is $p/N$), and then
evolve the system in time using only the total Hamiltonian $H$ given by Eq. 
(\ref{ham2}). Consider the current operator for the bond $(n,n+1)$,
\beq {\hat J}_n ~=~ -i ~(c_n^\da c_{n+1} ~-~ c_{n+1}^\da c_n), \eeq
or, in matrix form,
\beq ({\hat J}_n)_{jk} ~=~-i ~(\de_{j,n} \de_{k,n+1} ~-~\de_{j,n+1} \de_{k,n}).
\eeq
(Note that this matrix is Hermitian, imaginary and antisymmetric).
We can then obtain the average value of the current at that bond by 
calculating the expectation value of ${\hat J}_n$ over many time periods as 
described below; let us denote this average value by $I_{dc}$.
Note that in the absence of reservoirs, the system will not reach a steady 
state. The quantity $I_{dc}$ should therefore not be thought of a steady 
state current; it is merely the current averaged over many time periods. 
We will see that apart from the average part called $I_{dc}$, the current
continues to oscillate in an aperiodic manner even if we wait for a 
period of time which is much longer than $T$.

The time evolution of the system is governed by the unitary time evolution
operator
\beq U(t) ~=~ \lim_{M \to \infty} ~{\cal T} ~\prod_{j=1}^{M} ~e^{-i H (t_j) 
dt}, \label{ut} \eeq
where $\cal T$ stands for time ordering (namely, ${\cal T} {\cal O} (t_1)
{\cal O} (t_2) = {\cal O} (t_1) {\cal O} (t_2)$ if $t_1 > t_2$), and we have 
divided the time $t$ into $M$ equal steps, i.e., $dt=t/M$ and $t_j = (j-1/2)
dt$; eventually, we have to take the limit $M \to \infty$. We now consider 
$U(T)$ where $T$ is the time period. Let $v_j$ and $e^{i\ta_j}$ denote the 
eigenstates and corresponding eigenvalues of $U(T)$, with $j=1,2,\cdots,N$. 
Due to the periodicity of $H$ in time, we see that $U(sT) v_j = e^{is\ta_j} 
v_j$ for any positive integer $s$. Let us assume for simplicity that there 
is no degeneracy, so that $e^{i\ta_j} \ne e^{i\ta_k}$ if $j \ne k$; we can 
take the $v_j$'s to form an orthonormal basis. 

We observe that if the periodic potentials in Eq. (\ref{ham2}) are all
shifted in time by the same amount, i.e., $w_n (t) \to w_n (t + \tau)$, where
$\tau$ lies in the range $[0,T]$, this generally changes the unitary operator 
$U(T)$. However, if we redefine the vectors $v_j \to v'_j = U(\tau) v_j$, we 
can use the periodicity of $w_n (t)$ to show that the $v'_j$ will be 
eigenstates of the new operator $U'(T)$ with the {\it same} eigenvalues 
$e^{i\ta_j}$. This can be proved as follows. Let us introduce a two-parameter
notation for the time evolution operator
\beq U_2 (t',t) ~=~ \lim_{M \to \infty} ~{\cal T} ~\prod_{j=1}^{M} ~
e^{-i H (t_j) dt}, \label{ut2} \eeq
where $t_j = t + (j-1/2)dt$, and $dt=(t'-t)/M$. According to this 
notation, what we called
$U(t)$ earlier is actually $U_2 (t,0)$ and the time shifted operator $U'(T)$ 
is $U_2 (T+\tau, \tau)$. The periodicity of $w_n (t)$ then implies that 
$U_2 (T+\tau, \tau) U_2 (\tau,0) = U_2 (\tau,0) U_2(T,0)$, namely, that
$U'(T) U(\tau) = U(\tau) U(T)$. Hence $U(T) v_j = e^{i\ta_j} v_j$ implies
that $v'_j = U(\tau) v_j$ satisfies $U'(T) v'_j = e^{i\ta_j} v'_j$.
Next, let us define the matrix for the current averaged over one time period,
\beq ({\hat J}_n)_{jk} ~=~ \frac{1}{T} ~\int_0^T ~dt ~v_j^\da U^\da (t) 
{\hat J}_n U(t) v_k. \label{jnjk} \eeq
Then we can use the identity $U(t) U^\da (\tau) = U(t-\tau)$, for $t \ge \tau$,
to show that 
\beq v_j^\da U^\da (t) {\hat J}_n U(t) v_k ~=~ (v'_j)^\da U^\da (t-\tau) 
{\hat J}_n U(t-\tau) v'_k. \label{shift} \eeq
Integrating both sides of Eq. (\ref{shift}) over one time period $T$, we see 
that the quantities $({\hat J}_n)_{jk}$ defined in Eq. (\ref{jnjk}) are 
invariant under a shift in time by an arbitrary amount $\tau$.

We now begin with an initial one-particle state $\psi_a$ at time $t=0$. This 
can be written as a linear superposition, $\psi_a = \sum_j c_{aj} v_j$, where 
$c_{aj} = v_j^\da \psi_a$. If we evolve this initial state using the 
Hamiltonian $H$, the current averaged over an infinitely long time will be 
given by 
\bea J_n (\psi_a) &=& \lim_{s \to \infty} ~\frac{1}{sT} ~\int_0^{sT} ~dt ~
\psi_a^\da U^\da (t) {\hat J}_n U(t) \psi_a \non \\
&=& \lim_{s \to \infty} \frac{1}{s} ~\sum_{m=0}^{s-1} \sum_{j,k} ~
e^{im(\ta_k - \ta_j)} c_{aj}^* c_{ak} ({\hat J}_n)_{jk}. \non \\
& & \label{jpsi} \eea
Since we have assumed that there is no degeneracy, i.e., $e^{i(\ta_k - \ta_j)}
\ne 1$ if $j \ne k$, we see that
\beq \lim_{s \to \infty} ~\frac{1}{s} ~\sum_{m=0}^{s-1} e^{im(\ta_k - \ta_j)}
~=~ \de_{j,k}. \eeq
Hence Eq. (\ref{jpsi}) simplifies to
\beq J_n (\psi_a) ~=~ \sum_{j=1}^N ~|c_{aj}|^2 ~({\hat J}_n)_{jj}. \eeq
We thus have a simple expression for the current averaged over an infinite 
amount of time, even though the current will oscillate in an aperiodic manner 
at all times due to the factor of $e^{im(\ta_k - \ta_j)}$ in Eq. (\ref{jpsi}). 
Finally, if we start 
at $t=0$ with $p$ electrons occupying the orthonormal states $\psi_a$, where 
$a=1,2,\cdots,p$, the current averaged over an infinite time will be given by
\beq I_{dc} ~=~ \sum_{a=1}^p ~\sum_{j=1}^N ~|c_{aj}|^2 ~({\hat J}_n)_{jj}.
\label{idc} \eeq

We have numerically computed the diagonal part of the averaged current 
$({\hat J}_n)_{jj}$ for several different situations. For the case in which 
$H$ is real, we have made the following observations, all of which will be 
seen to be quite different from the situation on the infinite line.

\noi (i) If the periodic potentials $w_n (t)$ are shifted in time by an amount
$\tau$ as described above, the value of $({\hat J}_n)_{jj}$ changes in general. 
This is because the overlaps $|c_{aj}|^2$ of the new eigenstates $v'_j$ with 
the initial states $\psi_a$ is given by
\beq |c_{aj}|^2 ~=~ | v_j^\da U^\da (\tau) \psi_a |^2 \label{cajtau} \eeq
which will generally depend on $\tau$, except in the trivial limit in which
all the oscillating potentials $w_n (t)$ are set equal to zero. Hence the 
averaged current given in Eq. (\ref{idc}) will vary with $\tau$, even 
though $({\hat J}_n)_{jj}$ is independent of $\tau$ as discussed earlier. 
All this is quite different from the situation on the infinite line where 
the time averaged current does not change if all the periodic potentials
are shifted in time by the same amount. 

We note, however, that the dependence of $I_{dc}$ on the shift $\tau$ (or, 
equivalently, the overall phase of all the oscillating potentials) is weak 
if the amplitudes of all the oscillating potentials are small compared
to the hopping amplitude $\gamma$. This statement will be made more precise 
below.

\noi (ii) If the periodic potential is time reversal invariant, i.e., if $w_n
(t) = w_n (-t)$ which is also equal to $w_n (T-t)$ due to the periodicity in
time, then $({\hat J}_n)_{jj} = 0$ for each value of $j$. We will present an 
analytical proof of this below. From this it follows that $({\hat J}_n)_{jj} 
= 0$ even if the periodic potential is time reversal invariant only up to a 
shift, i.e., if $w_n (t) = w_n (\tau -t)$ for some value of $\tau$. This too 
differs from the situation on the infinite line where the averaged current is 
non-zero even if the periodic potential is time reversal invariant, as long 
as the static potentials break the left-right symmetry.

\noi (iii) If the periodic potential is applied to only one site, then
$({\hat J}_n)_{jj} = 0$ even if the potential has no particular symmetry in 
time. We have found this numerically for a wide variety of potentials $w_n 
(t)$; $({\hat J}_n)_{jj}$ always turns out to be zero to very high precision.
We have no analytical understanding of this remarkable result. This is also 
different from the situation on the infinite line where we have already seen 
that an oscillating potential applied to only one site can pump current if the
static potentials break left-right symmetry.

We will now prove analytically that $({\hat J}_n)_{jj} = 0$ if $H$ is real and
$w_n (t) = w_n (T-t)$ for all $n$. Following the notation of Eq. (\ref{ut}), 
this property of $w_n (t)$ implies that
\beq (e^{-iH(t)dt})^* ~=~ (e^{-iH(T-t)dt})^{-1}, \label{utsym1} \eeq
and therefore that $U^* (T) = U^{-1} (T)$. Hence $U(T) v_j = e^{i\ta_j} 
v_j$ implies that $U(T) v^*_j = e^{i\ta_j} v^*_j$. The non-degeneracy of the 
eigenvalues of $U(T)$ then implies that $v^*_j$ must be proportional to $v_j$.
Hence we can multiply the $v_j$'s by appropriate phases to ensure that $v^*_j 
= v_j$ for each value of $j$. Next, we can use Eq. (\ref{utsym1}) to show that 
\beq U(T-t) ~=~ U^* (t) U(T) \label{utsym2} \eeq
for any value of $t$ lying in the range $[0,T]$. 
We can now combine Eq. (\ref{utsym2}) with the reality of $v_j$ to 
show that $v_j^\da U^\da (T-t) {\hat J}_n U (T-t) v_j$, which must be real
since ${\hat J}_n$ is Hermitian, is equal to $v_j^\da U^\da (t) {\hat J}_n^* 
U (t) v_j = - v_j^\da U^\da (t) {\hat J}_n U (t) v_j$ since ${\hat J}_n^* = - 
{\hat J}_n$. We thus see that
\beq ({\hat J}_n)_{jj} ~=~ \frac{1}{T} ~\int_0^T ~dt ~v_j^\da U^\da (t) 
{\hat J}_n U(t) v_j ~=~ 0 \eeq
due to an exact cancellation of the integrand at the times $t$ and $T-t$.

\begin{figure}[htb]
\begin{center} \epsfig{figure=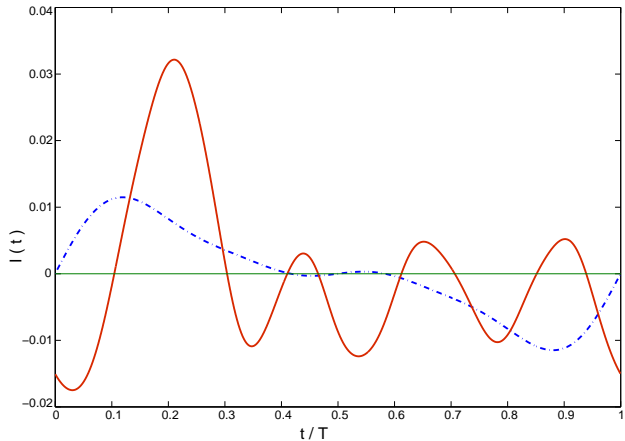,width=8.6cm} \end{center}
\caption{(Color online) Instantaneous current at various times over one time 
period $T = 10 \pi$ (corresponding to $\om = 0.2$) for an oscillating 
potential applied at one site, for the third eigenstate of $U(T)$ for a
six-site system. Two cases are considered: the oscillating potential is (i) 
time-reversal invariant (blue dashed line), and (ii) not time-reversal 
invariant (red solid line). A static potential of strength 1 is applied at site
2, while the oscillating potential is applied at site 3. For case (i) with 
an oscillating potential $\cos (\om t) + 0.2 \cos (2 \om t)$, we see an exact 
cancellation between the values at times $t$ and $T-t$, while case (ii) with 
an oscillating potential $\cos (\om t) + 0.5 \sin (2 \om t)$ does not show 
such a pair-wise cancellation. In both cases, the integrated current 
contributed by the third eigenstate is zero.} \end{figure}

In Fig. 4, we show that the current integrated over one time period is
zero if an oscillating potential is applied to only one site; as a specific
example, we have considered the contribution to the current from the third
eigenstate of $U(T)$ for a ring with six sites, when a static potential
and an oscillating potential are applied to two different sites. The
way in which the current integrates to zero is quite different depending on 
whether the oscillating potential is time reversal invariant or not.
If the oscillating potential is time reversal invariant, the integrated
current vanishes due to a pair-wise cancellation from the times $t$ and $T-t$
as we have proved above. But if the oscillating potential is not time reversal
invariant, there is apparently no simple symmetry reason for the vanishing
of the integrated current.

Since the integrated current vanishes if an oscillating potential is applied 
to only one site, we will now study what happens when oscillating potentials 
are applied to two sites. In all the figures discussed below (Figs. 5-9), we 
will consider a six-site system in which a static potential is applied at site
2, and oscillating potentials of the forms $b \cos (\om t + \phi_0)$ and $b 
\cos (\om t + \pi/2 + \phi_0)$ are applied at sites 3 and 4 respectively.

We first study the dependence of the pumped current on the overall phase 
$\phi_0$ of the oscillating potentials. We take $b=0.1$, and use Eqs. 
(\ref{idc}-\ref{cajtau}) to compute the averaged current as a function of 
$\phi_0 = 2 \pi \tau/T$.
In Fig. 5, we show the dependence of the current $I_{dc}$ on $\phi_0$ when the 
system has three electrons which corresponds to half-filling. We find that 
the variation of $I_{dc}$ from its mean value is about $8\%$. When we reduce 
$b$ by a factor of 2, we find that the mean value of $I_{dc}$ decreases by 
a factor of 4 while its variation with $\phi_0$ decreases by a factor of 
16. In the Appendix, we show that the mean value of $I_{dc}$ scales as 
$b^2$ which is in agreement with the 
numerical result quoted above. In addition, the numerics suggests that the
variation of $I_{dc}$ with $\phi_0$ scales as $b^4$. Thus, if the pumping 
amplitude $b$ is small compared to the hopping amplitude $\gamma$, the 
variation of the pumped current with $\phi_0$ becomes much smaller than 
the current itself.

\begin{figure}[htb]
\begin{center} \epsfig{figure=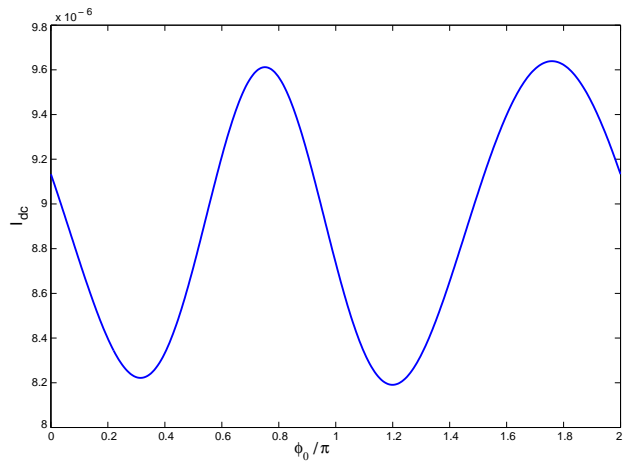,width=8.6cm} \end{center}
\caption{(Color online) Total pumped current versus the overall phase for
oscillating potentials applied at two sites, for a six-site ring with three 
electrons. A static potential of strength 1 is applied at site 2, while the 
oscillating potentials are applied at sites 3 and 4 with an amplitude 
$b = 0.1$, a frequency $\om = 0.2$, and a phase difference of $\pi/2$.} 
\end{figure}

Next, we consider the contributions to the averaged pumped currents,
$I_j \equiv ({\hat J}_n)_{jj}$ of the different eigenstates of $U(T)$, taking 
the overall phase to be given by $\phi_0 = 0$. For the six-site ring, with a 
static potential of strength 1 applied at site 2, the eigenvalues of the 
static part of the Hamiltonian, $H_0$, are equal to $-1.8912$, $-1.0000$, 
$-0.7046$, $1.0000$, $1.3174$, and $2.2784$. In Fig. 6, we show the six 
contributions $I_j$ as a function of the oscillation amplitude $b$, when 
the oscillation frequency $\om = 0.2$ is non-resonant, i.e., it does not 
correspond to the difference between any two energy levels of $H_0$. We see 
that all the curves show a quadratic dependence on $b$, although two of them 
lie very close to zero; the quadratic dependence is in accordance with the 
results derived in the Appendix for the non-resonant case. In Fig. 7, we show 
the total pumped current $I_{dc}$ versus $b$ when there are three electrons, 
where $I_{dc}$ has been calculated using Eq. (\ref{idc}) and the individual 
pumped currents shown in Fig. 6.

\begin{figure}[htb]
\begin{center} \epsfig{figure=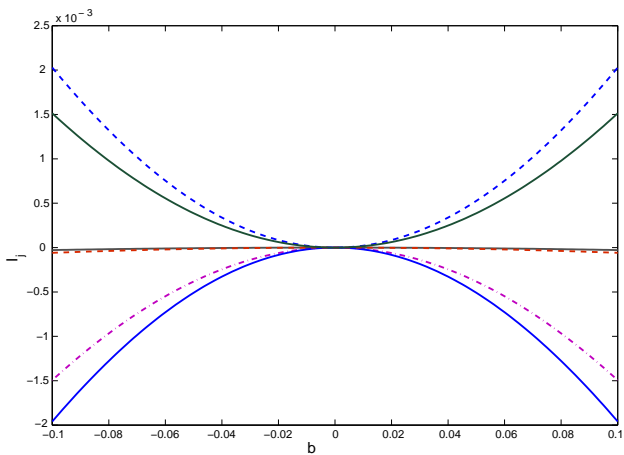,width=8.6cm} \end{center}
\caption{(Color online) Pumped currents for all the eigenstates of $U(T)$ for 
oscillating potentials applied at two sites versus the oscillation amplitude 
$b$, for a six-site ring. A static potential of strength 1 is applied at site 
2, while the oscillating potentials are applied at sites 3 and 4 with a 
phase difference of $\pi/2$ and a non-resonant frequency $\om = 0.2$. In all 
cases, the current is proportional to $b^2$ for small $b$.} \end{figure}

\begin{figure}[htb]
\begin{center} \epsfig{figure=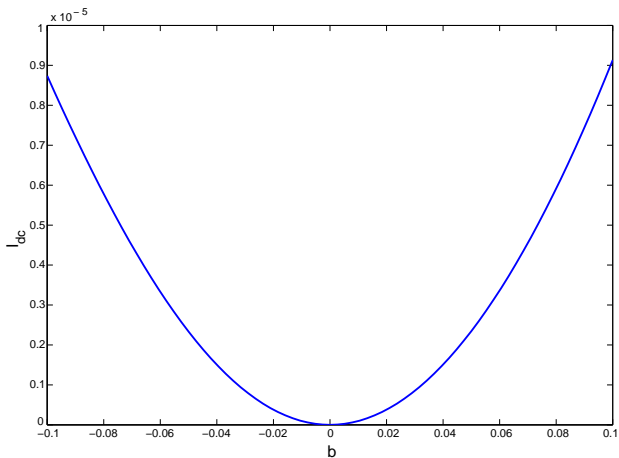,width=8.6cm} \end{center}
\caption{(Color online) Total pumped current for oscillating potentials 
applied at two sites versus the oscillation amplitude $b$, for a six-site 
ring with three electrons. All the parameters are the same as in Fig. 6.} 
\end{figure}

In Fig. 8, we show the contributions $I_j$ 
as a function of $b$, when the oscillation frequency $\om = 1.7046$ is 
resonant; it corresponds to the difference between the third and fourth 
energy levels of $H_0$. We see that four of the curves lie very close 
to zero and show a quadratic dependence on $b$, but the other two curves, 
which have large contributions from the third and fourth eigenstates of $H_0$,
show a linear dependence on $b$; the linear dependence agrees with the results
derived in the Appendix for the resonant case. In Fig. 9, we show the total 
current $I_{dc}$ versus $b$ when there are three electrons. This current 
varies quadratically with $b$ for the following reasons. First, the two 
states which show a linear dependence on $b$ in Fig. 8 are found to contribute
with equal weight to $I_{dc}$. Secondly, the slopes of these two curves in 
Fig. 8 are equal and opposite at $b=0$. Hence the linear dependences on $b$ 
of these two curves cancel out when they are added up using Eq. (\ref{idc})
to calculate $I_{dc}$ which therefore shows a quadratic dependence on $b$. We 
also observe that the currents in Fig. 9 are about 90 times those in Fig. 7,
for the same values of $b$; this shows the effectiveness of charge pumping 
at resonant versus non-resonant frequencies.

\begin{figure}[htb]
\begin{center} \epsfig{figure=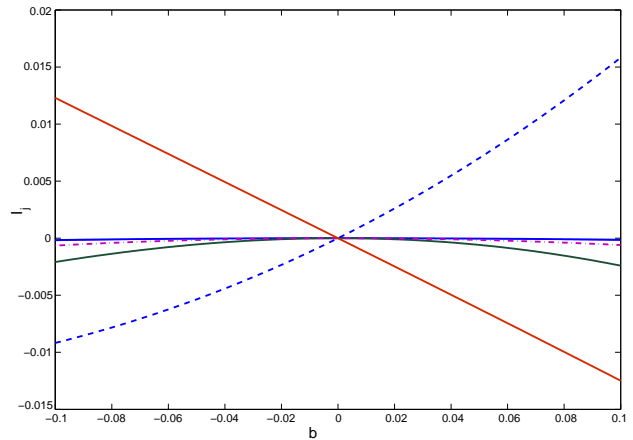,width=8.6cm} \end{center}
\caption{(Color online) Pumped currents for all the eigenstates of $U(T)$ for 
oscillating potentials applied at two sites versus the oscillation 
amplitude $b$, for a six-site ring. A static potential of strength 1 is 
applied at site 2, while the oscillating potentials are applied at sites 3 
and 4 with a phase difference of $\pi/2$ and a resonant frequency $\om = 
1.7046$. In four cases, the current is proportional to $b^2$, but in two 
cases, the current is proportional to $b$ for small $b$ (see text for 
details).} \end{figure}

\begin{figure}[htb]
\begin{center} \epsfig{figure=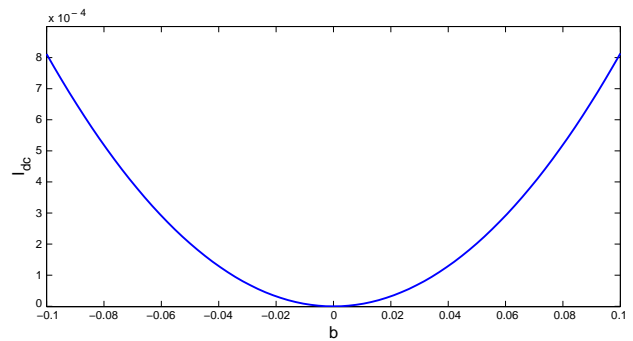,width=8.6cm} \end{center}
\caption{(Color online) Total pumped current for oscillating potentials 
applied at two sites versus the oscillation amplitude $b$, for a six-site 
ring with three electrons. All the parameters are the same as in Fig. 8.} 
\end{figure}

Finally, let us briefly discuss what happens if the Hamiltonian is not
real. For instance, if there is a magnetic flux passing through the ring, 
its effects can be studied by making the hopping amplitudes $\ga$ complex
in Eq. (\ref{ham2}). In such cases, we find that there can be a net dc 
current in the ground state even in the absence of any oscillating potentials;
this is called a persistent current. The value of the dc current can then 
change even if we apply oscillating potentials which are time reversal 
invariant or if an oscillating potential is applied at only one site.

\section{Discussion}

To summarize, we have studied charge pumping both on an infinite line and on 
a finite ring, for a system of non-interacting electrons at zero temperature.
For the infinite line with reservoirs at the same chemical potential on the 
two sides, we have verified, in agreement with earlier work, that oscillating 
potentials applied to one or two sites can pump a net dc current as long as 
the left-right symmetry is broken by some static potentials; the oscillating 
potentials do not need to break time reversal invariance \cite{moskalets1}. 
As a specific example, we have studied pumping in a situation where the Fermi 
energy differs from the resonant energy, but the pumping frequency is equal to 
the difference of the those two energies. We have contrasted the cases
of pumping at one site and two sites. We have also shown 
that the $\sin \phi$ rule, which holds in the adiabatic limit ($\om \to 0$) 
for pumping by two oscillating potentials with a phase difference of $\phi$, 
fails if $\om$ is larger than the resonance width.

For a purely Hamiltonian evolution on a finite ring, we have shown that if 
both the static and time-dependent parts of the Hamiltonian are real, the net 
dc current is zero either if the oscillating potentials are time reversal 
invariant or if the oscillating potential is 
applied to only one site. We have presented an analytical proof of the former 
statement and have presented numerical evidence for the latter. It would be 
very useful if an analytical proof could be found for the vanishing of the 
current for one-site pumping with an arbitrary time dependence.

We have argued that for a purely Hamiltonian evolution, the system on a ring 
does not reach a steady state; the current averaged over one time period $T$ 
continues to vary with time in an aperiodic manner although there is a simple 
expression for the current averaged over an infinitely long time. Further, 
the averaged current depends on the initial state and on the overall phase 
of the oscillating potentials since these two determine the overlap between 
the initial state and the eigenstates of the evolution operator $U(T)$.

The facts that a steady state is not reached and that breaking of time 
reversal invariance of the oscillating potentials is required for charge 
pumping to occur on a finite ring but not on an infinite line is due to an 
important difference between the models that we have assumed in the two cases. 
For the finite ring, we have assumed a purely Hamiltonian evolution, with no 
mechanisms for momentum relaxation or phase decoherence. On the other hand, 
the study of the infinite line assumed that there are reservoirs which are 
maintained at certain chemical potentials. This implicitly assumes that there 
are mechanisms for energy or momentum relaxation; for instance, if an 
electron emerges from the double barrier with an energy equal to the resonant 
energy $E_r$, we assume that when it reaches one of the reservoirs, it will 
relax down to the Fermi energy $E_F$ if $E_r > E_F$. It appears that such 
relaxation processes, which we have not explicitly referred to in our 
calculations but which are necessarily present in the reservoirs, effectively
lead to a breaking of time reversal invariance which is required to have a 
net dc current on the infinite line. An earlier study has shown that if a 
finite ring is coupled to reservoirs, then charge pumping can occur even if 
the oscillating potentials are time reversal invariant \cite{arrachea2}.

It is also worth noting that on a finite ring, an electron which emerges to 
the right from the region containing the potentials (both static and 
oscillating) eventually comes back and enters the same region from the left. 
Thus the outgoing current on the right of that region must be equal to the
incoming current on the left, when these currents are averaged over a long
time. Similarly, the averaged outgoing current on the left of the region must 
be equal to the averaged incoming current on the right. These relations do 
not hold on the infinite line since an electron going out to the right (left) 
does not return to the left (right) of the region containing the potentials.

We now turn to the possible experimental implications of our results. It is 
possible that the dependence of the averaged pumped current on the initial 
state and on the overall phase of the oscillating potentials on a ring with 
no mechanisms for momentum and phase relaxation can be observed experimentally.
For mesoscopic rings which have a large number of impurities which lead to 
elastic scattering, the transport is diffusive and is characterized by a 
diffusive round-trip time $\tau_d$ which is equal to $3L^2/(v_F l_c)$,
where $l_c$ is the mean free path and $L$ is the circumference of the ring
\ct{bluhm}. Another important length scale for ring systems is the phase 
coherence length $l_\phi$ which depends strongly on the temperature 
\ct{kane,jariwala,deblock}. However, one may consider wires which have a 
sufficiently low density of scatterers and are at sufficiently low 
temperatures that both $l_c$ and $l_\phi$ are much larger $L$. In such a 
situation, we expect that if the pumping potentials are suddenly switched 
on, then the averaged pumped current will
initially depend on the overall phase of the potentials. Eventually, at time 
scales which are much longer than $l_\phi/v_F$ and $\tau_d$, the averaged 
pumped current will no longer remember the overall phase; however, a
proof of this is beyond the scope of our analysis since we have not 
introduced any mechanisms for momentum and phase relaxation. 
 
Another arena where our results could possibly be tested is the field of 
molecular electronics \ct{molen,ander}. We would like to propose an 
experimental set-up as follows. The conductance properties of aromatic 
molecules (which typically have a ring-like structure) have been studied 
extensively for several years. Typically, the transport properties of such 
molecules are studied by depositing the molecule on a substrate and using an 
scanning tunnel microscope (STM) 
tip from above to probe the molecule \ct{molen}. The number of electrons 
in the molecule can be fixed initially by bringing the STM tip close to the 
molecule and applying the correct potential to the tip. We propose that the 
distance of the STM tip from the molecule can then be increased so that 
electron tunneling between the STM and the molecule becomes negligible; 
however the STM tip can still be used to induce an on-site pumping 
(oscillating) potential at a particular atom. This can be used to study
the effect of pumping on the electronic transport through the molecule.
A theoretical analysis of this would also require us to consider the
effect of interactions between the electrons.

An interesting problem for future studies may be to include the effects of 
interactions between the electrons \ct{riwar}. In particular, one can study 
whether interactions modify some of the peculiar features observed on the 
finite ring, such as the absence of pumping for an oscillating potential 
which is applied to only one site. In this context, we note that interactions
between electrons are believed to play a role in determining the magnitude 
of the persistent current in rings placed in a magnetic field \ct{imry}.

\section*{Acknowledgments}

We thank A. Agarwal, A. Dhar, D. Dhar, A. Jayannavar, M. Moskalets and 
M. Stone for stimulating discussions. A. S. thanks CSIR, India for financial 
support, and D. S. thanks DST, India for financial support under Project No. 
SR/S2/CMP-27/2006.

\vskip .8 true cm
\centerline{\bf Appendix}
\vskip .4 true cm

We will use first order perturbation theory to derive expressions for the 
eigenstates of $U(T)$ on a finite ring and the integrated current in those 
eigenstates, for both the non-resonant and 
resonant cases. The treatment below will be seen to have interesting 
parallels with the use of first order perturbation theory to obtain the 
eigenstates of a time-independent Hamiltonian in the non-degenerate and 
degenerate cases respectively.

The Hamiltonian of interest is $H=H_0 + V(t)$, where $V(t)$ has a
periodicity given by $T = 2\pi/\om$. Let us first set $V =0$.
$H_0$ will have a complete set of orthonormal 
eigenstates $\psi_j$ and eigenvalues $E_j$, where we will assume, for
simplicity, that the $E_j$'s are non-degenerate. We will also assume that
$H_0$ is real; hence the $\psi_j$'s can be chosen to be real. If a unitary
evolution operator $U_0 (T)$ is constructed using $H_0$, its eigenstates
and eigenvalues will be given by $\psi_j$ and $e^{-iE_j T}$. Note that since
the wave functions $\psi_j$ are real and ${\hat J}_n^* = {\hat J}_n^T = - 
{\hat J}_n$, we have the useful relations
\bea \psi_j^\da {\hat J}_n \psi_j &=& 0, \non \\
\psi_j^\da {\hat J}_n \psi_k &=& - ~ \psi_k^\da {\hat J}_n \psi_j ~~~{\rm 
for}~~~ j \ne k. \label{jsym} \eea

We now turn on the time-dependent perturbation $V$ which will be assumed 
to satisfy Eq. (\ref{vzero}). There are two different possibilities which we 
will discuss separately: 

\noi (i) Non-resonant case where $E_j - E_k$ is not an integer multiple of 
$\om$, i.e., $e^{-i(E_j-E_k) T} \ne 1$, for any pair of states $j, k$, and 

\noi (ii) Resonant case where $E_j - E_k$ is an integer multiple of $\om$, 
i.e., $e^{-i(E_j-E_k) T} = 1$, but $E_j \ne E_k$, for some pair of states 
$j, k$.

\noi (iii) Resonant case where $E_j = E_k$ for some pair of states $j, k$.

\vskip .4 true cm
\centerline{\bf Non-resonant case}
\vskip .4 true cm

We will find the eigenstate of $U(T)$, called $v_1$, which differs at first 
order in $V$ from a particular eigenstate of $U_0 (T)$, say, $\psi_1$. 
We assume that $v_1 (t) = U(t) v_1$ has an expansion of the form 
\beq v_1 (t) ~=~ \sum_{j=1}^N ~c_j (t) ~e^{-iE_j t} ~\psi_j, \label{v1t} \eeq
where we choose $c_1 (0) = 1$. (We are not worrying about the normalization
of $v_1$ here, although we see that $v_1$ is normalized to 1 up to zero-th
order in $V$). We expect that the deviation of $c_1 (t)$ from
1 at different values of $t$ and also the values of $c_j (t)$ for $j \ne 1$
will be of order $V$. From the Schr\"odinger equation 
\beq i\frac{dv_1 (t)}{dt} ~=~ [H_0 ~+~ V(t)] ~v_1 (t), \label{sch} \eeq
we find the following equations to first order in $V$,
\bea i\frac{dc_1(t)}{dt} &=& c_1 (t) ~\la \psi_1 | V(t) | \psi_1 \ra,
\non \\
i\frac{dc_j(t)}{dt} &=& c_1 (t) ~\la \psi_j | V(t) | \psi_1 \ra ~
e^{i(E_j - E_1)t} \label{cjt1} \eea
for $j \ne 1$. At first order, we can replace $c_1(t)$ by $c_1(0)=1$ in 
Eqs. (\ref{cjt1}). This gives the solution
\bea c_1(t) &=& 1 ~-~ i \int_0^t ~dt' \la \psi_1 | V(t') | \psi_1 \ra, \non \\
c_j(t) &=& \al_j - ~i \int_0^t ~dt' \la \psi_j | V(t') | \psi_1 \ra ~
e^{i(E_j - E_1)t'}, \label{cjt2} \eea
for $j \ne 1$, where $\al_j$ are constants of integration which can be fixed 
as follows. Since $c_1 (T) = c_1 (0)$ due to Eq. ({\ref{vzero}), we see from 
the $\psi_1$ term in Eq. (\ref{v1t}) that $v_1 (T) = e^{-iE_1 T} v_1(0)$, i.e.,
$v_1$ is an eigenstate of $U(T)$ with eigenvalue $e^{-iE_1 T}$. We therefore
demand that this should also be true for all the other terms $\psi_j$ for
$j \ne 1$ in Eq. (\ref{v1t}). We therefore require that $c_j (T) e^{-iE_j T}
= e^{-iE_1 T} c_j (0)$ for all $j \ne 1$. This fixes the value of the 
constants $\al_j$ in Eq. (\ref{cjt2}), and we find that 
\bea c_j(t) &=& - i~\frac{\int_0^T ~dt' \la \psi_j | V(t') | \psi_1 \ra ~
e^{i(E_j - E_1)t'}}{e^{i(E_j - E_1)T} ~-~ 1} \non \\
& & -i~ \int_0^t ~dt' \la \psi_j | V(t') | \psi_1 \ra ~e^{i(E_j - E_1)t'}
\label{cjt3} \eea
for $j \ne 1$. Combining Eqs. (\ref{v1t}), (\ref{cjt2}) and (\ref{cjt3}),
we have an expression for the eigenstate of $U(T)$ to first order in
$V$. Note that the corresponding eigenvalue remains $e^{-iE_1 T}$ to
this order; this is a consequence of the choice made in Eq. (\ref{vzero}).

We can now calculate the current averaged over one time period for one of
these eigenstates, 
\beq ({\hat J}_n)_{jj} ~=~ \frac{1}{T} ~\int_0^T ~dt ~v_j^\da U^\da (t) 
{\hat J}_n U(t) v_j. \label{jnjj} \eeq
Due to the fact that $\psi_j^\da {\hat J}_n \psi_j = 0$, the current can only 
get a contribution from cross-terms of the form $\psi_j^\da {\hat J}_n 
\psi_k$ for $j \ne k$ arising from Eq. (\ref{v1t}). From Eqs. 
(\ref{cjt2}-\ref{cjt3}), we see that $({\hat J}_n)_{jj}$ has no contributions
of order 1, while the contributions of order $V$ can only come from a 
cross-term between $\psi_1$ and $\psi_j$ for $j \ne 1$. Such contributions
are proportional to
\beq \int_0^T ~dt ~e^{i(E_1 - E_j)t} ~c_j (t), \label{intcj} \eeq
where $c_j (t)$ is given in Eq. (\ref{cjt3}). We can now do the integral in 
Eq. (\ref{intcj}) explicitly and we find that it vanishes. We therefore 
conclude that $({\hat J}_n)_{jj}$ only receives contributions of second 
order and higher in $V$.

\vskip .4 true cm
\centerline{\bf Resonant case with $E_1 \ne E_2$}
\vskip .4 true cm

Let us now consider the case when two eigenstates of $U_0 (T)$, say, $\psi_1$ 
and $\psi_2$, have the same eigenvalue $e^{-iE_1T} = e^{-iE_2T}$ which 
implies that $E_1 - E_2$ is an integer multiple of $\om$. We will assume, 
however, that $E_1 \ne E_2$. We will consider only the states 1 and 2,
and will study how they can be combined to form eigenstates of $U (T)$
to first order in $V$. Let us consider an expansion of the form
\beq v (t) ~=~ c_1 (t) ~e^{-iE_1 t} ~\psi_1 ~+~ c_2 (t) ~e^{-iE_2 t} ~
\psi_2, \label{vt} \eeq
where we now assume that both $c_1 (t)$ and $c_2 (t)$ are of order 1. (This
is in contrast to the non-resonant case where only $c_1 (t)$ was taken to 
be order 1). Further, let us take $c_1 (0) = c_{10}$ and $c_2 (0) = c_{20}$;
we will assume that the deviations of $c_1 (t)$ and $c_2 (t)$ from $c_{10}$ 
and $c_{20}$ respectively will be of order $V$ at all values of $t$.
The Schr\"odinger equation in Eq. (\ref{sch}) now gives
\bea i\frac{dc_1(t)}{dt} &=& c_1 (t) ~\la \psi_1 | V(t) | \psi_1 \ra 
\non \\
& & +~ c_2 (t) e^{i(E_1 - E_2)t} ~\la \psi_1 | V(t) | \psi_2 \ra, \non \\
i\frac{dc_2(t)}{dt} &=& c_2 (t) ~\la \psi_2 | V(t) | \psi_2 \ra \non \\
& & +~ c_1 (t) e^{i(E_2 - E_1)t} ~\la \psi_2 | V(t) | \psi_1 \ra. 
\label{c12t1} \eea
At first order, we can replace $c_1(t)$ and $c_2 (t)$ by $c_{10}$ and
$c_{20}$ on the right hand sides in Eqs. (\ref{c12t1}). This gives
\bea c_1 (t) &=& c_{10} ~-~ i ~c_{10} ~\int_0^t ~dt' \la \psi_1 | V(t') | 
\psi_1 \ra \non \\
& & -~ i ~c_{20} ~\int_0^t ~dt' \la \psi_1 | V(t') | \psi_2 \ra ~
e^{i(E_1 - E_2)t'}, \non \\
c_2 (t) &=& c_{20} ~-~ i ~c_{20} ~\int_0^t ~dt' \la \psi_2 | V(t') | 
\psi_2 \ra \non \\
& & -~ i ~c_{10} ~\int_0^t ~dt' \la \psi_2 | V(t') | \psi_1 \ra ~
e^{i(E_2 - E_1)t'}. \eea
Using Eq. (\ref{vzero}), we see that
\bea c_1 (T) &=& c_{10} ~-~ i~c_{20} ~\int_0^T ~dt \la \psi_1 | V(t) | 
\psi_2 \ra ~e^{i(E_1 - E_2)t}, \non \\
c_2 (T) &=& c_{20} ~-~ i~c_{10} ~\int_0^T ~dt \la \psi_2 | V(t) | 
\psi_1 \ra ~e^{i(E_2 - E_1)t}. \non \\ \label{c12t2} \eea
If we now demand that the state in Eq. (\ref{vt}) satisfies $v(T) = e^{i\ta}
v(0)$, i.e., that $v(0)$ is an eigenstate of $U(T)$ with eigenvalue 
$e^{i\ta}$, we find that 
\beq \frac{c_1 (T) ~e^{-iE_1 T}}{c_1 (0)} ~=~ \frac{c_2 (T) ~e^{-iE_2 T}}{
c_2 (0)} ~=~ e^{i\ta}. \label{eit} \eeq
This implies that 
\beq \left( \frac{c_{20}}{c_{10}} \right)^2 ~=~ \frac{\int_0^T ~dt \la 
\psi_2 | V(t) | \psi_1 \ra ~e^{i(E_2 - E_1)t}}{\int_0^T ~dt \la \psi_1 | 
V(t) | \psi_2 \ra ~e^{i(E_1 - E_2)t}}. \eeq
Defining 
\beq \al ~=~ \frac{1}{T} ~\int_0^T ~dt \la \psi_1 | V(t) | \psi_2 \ra ~
e^{i(E_1 - E_2)t}, \eeq
we see that
\beq \frac{c_{20}}{c_{10}} ~=~ \pm \sqrt{\frac{\al^*}{\al}}, \label{pm} \eeq
so that $c_{20}/c_{10}$ is a pure phase. Using Eqs. (\ref{c12t2}-\ref{eit}), 
we see that the $\pm$ sign in Eq. (\ref{pm}) corresponds to two solutions 
$v_{\pm}(t)$ which satisfy $U(T) v_{\pm} (0) = e^{i\ta_{\pm}} v_{\pm} (0)$, 
where 
\bea e^{i\ta_{\pm}} &=& e^{-iE_1 T} ~\left[ 1 ~-~ i \al T~\frac{c_{20}}{c_{10}}
\right] \non \\
&=& e^{-iT ~(E_1 \pm |\al|)} \eea
to first order in $\al$. We thus see that the degeneracy of eigenvalues of
$U_0 (T)$ is broken at first order in $V$, with the phases $\ta_{\pm}$
being split by equal and opposite amounts.

Finally, we can compute the current averaged over one time period as defined 
in Eq. (\ref{jnjj}) for either one of the states, $v_+$ or $v_-$. Once again, 
only cross-terms of the form $\psi_1^\da {\hat J}_n \psi_2$ will contribute. 
We find that terms of order 1 vanish because they are of the form
\beq \int_0^T dt [c_{10}^* c_{20} e^{i(E_1 - E_2)t} \psi_1^\da {\hat J}_n 
\psi_2 + c_{20}^* c_{10} e^{i(E_2 - E_1)t} \psi_2^\da {\hat J}_n \psi_1 ], \eeq
and $\int_0^T ~dt~ e^{i(E_1 - E_2)t} = 0$ because $E_1 - E_2$ is an integer
multiple of $\om$ but $E_1 \ne E_2$. However, unlike the non-degenerate case, 
there is now no reason for contributions of order $V$ to vanish in general. 
Hence $({\hat J}_n)_{jj}$ can get contributions at first order in $V$.

\vskip .4 true cm
\centerline{\bf Resonant case with $E_1 = E_2$}
\vskip .4 true cm

Finally, let us consider the case when two eigenstates of $H_0$, 
say, $\psi_1$ and $\psi_2$, have $E_1 = E_2$. This implies that they
also have the same eigenvalue $e^{-iE_1T} = e^{-iE_2T}$ of $U_0 (T)$. 
As in the previous subsection, we will study how these two states can be 
combined to form eigenstates of $U (T)$ to first order in $V$.

We can check that all the discussion from Eq. (\ref{vt}) to (\ref{c12t2})
will remain valid in this case, except that we have to substitute $E_1 = E_2$
everywhere. We then find that $c_1 (T) = c_1 (0) = c_{10}$ and $c_2 (T) = 
c_2 (0) = c_{20}$ due to Eq. (\ref{vzero}). Further,
there is now no relation between $c_{10}$ and $c_{20}$; we can 
choose $c_{10}$ and $c_{20}$ in an arbitrary way to obtain two orthonormal 
states $v_{\pm}$ which are eigenstates of $U(T)$ with the same eigenvalue 
$e^{-iE_1T}$. Thus the eigenvalues of $U(T)$ and their degeneracy do not 
change to first order in $V$.

When we compute the current averaged over one time period as defined 
in Eq. (\ref{jnjj}) for either one of the states, $v_+$ or $v_-$, we
find that, depending on the choices of $c_{10}$ and $c_{20}$, we can get 
a contribution to zero-th order in $V$. To be explicit, this is given by
\bea & & \int_0^T ~dt ~[c_{10}^* c_{20} ~\psi_1^\da {\hat J}_n \psi_2 ~+~ 
c_{20}^* c_{10} ~\psi_2^\da {\hat J}_n \psi_1 ] \non \\
& & = T ~(c_{10}^* c_{20} ~-~ c_{20}^* c_{10}) ~\psi_1^\da {\hat J}_n 
\psi_2, \eea
where we have used Eq. (\ref{jsym}).

It may seem strange that if $E_1 = E_2$, there can be a non-zero current in 
the states $v_+$ and $v_-$ separately even at the zero-th order in $V$. 
However, we note that an energy degeneracy 
usually does not occur if static potentials are present. But if there are
no static potentials present, then the system is translation invariant,
and we see that there are momentum eigenstates (with momenta $\pm k$)
which carry equal and opposite currents even in the limit that the 
amplitudes of the oscillating potentials go to zero.

\end{document}